\documentclass[draft,pdf]{aipproc}

\usepackage{graphicx}

\layoutstyle{6x9}

\date{}

\begin{document}

\newcommand{\be}{\begin{eqnarray}}
\newcommand{\ee}{\end{eqnarray}}


\baselineskip 18pt
\title{Random Matrices}


\author
{M.A.~STEPHANOV$^1$, J.J.M.~VERBAARSCHOT$^2$, 
and T.~WETTIG$^3$}{
address={ $^1$ Institute for Theoretical Physics, SUNY, Stony Brook, 
	NY 11794-3840\\
	$^2$ Department of Physics and Astronomy, SUNY, Stony Brook, NY
  	11794-3800\\ 
  	$^3$Institut f\"ur Theoretische Physik, Technische Universit\"at
  	M\"unchen, D-85747 Garching, Germany} }

\begin{abstract} {We review elementary properties of random matrices and 
discuss widely used mathematical methods for both
hermitian and nonhermitian random matrix ensembles. Applications to
a wide range of physics problems are summarized. 
This paper originally appeared as an article in the Wiley
Encyclopedia of Electrical and Electronics Engineering.}
\end{abstract}

\vfill
\maketitle
\newpage

\section*{Introduction}

In general, {\em random matrices} are matrices whose matrix elements
are {\em stochastic variables}.  The main goal of {\em Random Matrix
Theory} ({\em RMT}) is to calculate the statistical properties of
eigenvalues for very large matrices which are important in many
applications.  Ensembles of random matrices first entered 
in the mathematics literature as a $p$-dimensional generalization of
the $\chi^2$-distribution \cite{Wishart}. Ensembles of real symmetric
random matrices with independently distributed Gaussian matrix elements
were introduced in the physics literature in order to describe the spacing
distribution of nuclear levels \cite{Wigner}.  The theory of 
 {\em Hermitian} random matrices was first worked out in a series of
seminal papers by Dyson \cite{Dyson}. Since then, RMT has had
applications in many different branches of physics ranging from sound
waves in aluminum blocks to quantum gravity.  For an overview of the
early history of RMT we refer to the book by Porter \cite{Porter}.  An
authoritative source on RMT is the book by Mehta \cite{Meht91}.  For a
comprehensive review including the most recent developments we refer
to Ref.~\cite{Guhr98}.

Generally speaking, random matrix ensembles provide a statistical description
of a complex interacting  system. Depending on the hermiticity
properties of the interactions, one can distinguish 
two essentially different classes of random
matrices: Hermitian matrices with real eigenvalues and matrices
without hermiticity properties with eigenvalues scattered in the
complex plane. We will first give an overview of the ten different
classes of Hermitian random matrices and then briefly discuss
non-Hermitian random matrix ensembles. 

The best known random matrix ensembles are the {\em Wigner-Dyson ensembles}
which are ensembles of Hermitian matrices with matrix elements
distributed according to
\be
P(H)DH = {\cal N}e^{-\frac {N\beta}{4} {\rm Tr} H^\dagger H} DH. 
\label{probwd}
\ee
Here, $H$ is a Hermitian  $N\times N$ matrix with {\em real}, {\em complex}, or
{\em quaternion real} matrix elements. The corresponding random matrix
ensemble is characterized by the {\em Dyson index} $\beta =1,$ 2, and
4, respectively.  The measure $DH$ is the {\em Haar measure} which is
given by the product over the independent differentials.  The
normalization constant of the probability distribution is denoted by
${\cal N}$.  The probability distribution (\ref{probwd}) is invariant
under the transformation
\be H \rightarrow
U H U^{-1},
\label{sim}
\ee
where $U$ is an {\em orthogonal} matrix  for $\beta = 1$, a {\em
  unitary} matrix for $\beta = 2$, and a {\em symplectic} matrix for
$\beta = 4$.  This is the reason why these ensembles are known as the
{\em Gaussian Orthogonal Ensemble (GOE)}, the {\em Gaussian Unitary
  Ensemble (GUE)}, and the {\em Gaussian Symplectic Ensemble (GSE)},
respectively. The GOE is also known as the {\em Wishart distribution}.
Since both the eigenvalues of $H$ and the Haar measure $DH$ are
invariant with respect to (\ref{sim}), the eigenvectors and the
eigenvalues are independent with the distribution of the eigenvectors
given by the the invariant measure of the corresponding orthogonal,
unitary, or symplectic group.

There are two ways of arriving at the probability distribution
(\ref{probwd}).  First, from the requirement that the matrix elements
are independent and are distributed with the same average and variance
for an ensemble invariant under (\ref{sim}).  Second, by requiring
that the probability distribution maximizes the {\em information
  entropy} subject to the constraint that the average and the variance
of the matrix elements are fixed.

A second class of random matrices are the {\em chiral ensembles}
\cite{Verb94a} with the chiral symmetries of the {\em QCD Dirac
  operator}.  They are defined as the ensembles of $N\times N$
Hermitian matrices with block structure
\be\label{hc}
H= \left ( \begin{array}{cc} 0 & C \\ C^\dagger & 0 \end{array} \right )
\label{chRMT}
\ee
and probability distribution given by 
\be
P(C)DC = {\cal N}{\det}^{N_f} 
\left ( \begin{array}{cc} 0 & C \\ C^\dagger & 0 \end{array} \right )
e^{-\frac {N\beta}{4} {\rm Tr} C^\dagger C}
DC. 
\label{chRMTprob}
\ee
Again, $DC$ is the Haar measure, and $N_f$ is a real parameter
(corresponding to the number of quark flavors in QCD).  
The matrix $C$ is a rectangular $n \times (n+\nu)$ matrix.
Generically, the matrix $H$ in (\ref{hc})
 has exactly $|\nu|$ zero eigenvalues. Also generically,  
the QCD Dirac operator corresponding to a field configuration
with the {\em topological charge} $\nu$ has exactly $|\nu|$ zero eigenvalues,
in accordance with the {\em Atiyah-Singer index theorem}.  
For this reason, $\nu$ is identified as the topological quantum
number. The normalization constant of the probability distribution is
denoted by ${\cal N}$.  Also in this case one can distinguish
ensembles with real, complex, or quaternion real matrix elements. They
are denoted by $\beta= 1$, $\beta = 2$, and $\beta = 4$, respectively.
The invariance property of the chiral ensembles is given by
\be
C \rightarrow U C V^{-1},
\ee
where $U$ and $V$ are orthogonal, unitary, and symplectic matrices,
respectively. For this reason, the corresponding ensembles are known
as the {\em chiral Gaussian Orthogonal Ensemble (chGOE)}, the {\em 
  chiral Gaussian Unitary Ensemble (chGUE)}, and the {\em chiral
  Gaussian Symplectic Ensemble (chGSE)}, respectively. A two sub-lattice
model with diagonal disorder in the chGUE class was first considered 
in \cite{Gade}.

A third class of random matrix ensembles occurs in the description of
{\em  disordered superconductors}. Such ensembles with the symmetries
of the {\em  Bogoliubov-de Gennes Hamiltonian} have the block structure
\be
H =\left ( \begin{array}{cc}  A & B \\ B^\dagger & -A^T \end{array} \right),
\ee
where $A$ is Hermitian and, depending on the underlying symmetries,
the matrix $B$ is symmetric or anti-symmetric. The probability
distribution is given by
\be
P(H)DH = {\cal N}\exp 
\left (-\frac {N\beta}{4} {\rm Tr} H^\dagger H\right ) DH,
\label{prob}
\ee
where $DH$ is the Haar measure and ${\cal N}$ is a normalization
constant.  For symmetric $B$ the matrix elements of $H$ can be either
complex (C) or real (CI). For anti-symmetric $B$ the matrix elements
of $H$ can be either complex (D) or quaternion real (DIII).  The name
of the ensembles (in parentheses) refers to the {\em  symmetric
  space} to which they are tangent to. Since they were first
introduced by Altland and Zirnbauer \cite{AZ,class} we will call them
the {\em  Altland-Zirnbauer ensembles}. A hopping model based on the
class CI first entered in \cite{Oppermann}.

A key ingredient in the classification of a Hamiltonian in terms of
one of the above random matrix ensembles is its {\em  anti-unitary
  symmetries}.  An {\em  anti-unitary operator} can be written as
\be
U = A K,
\ee
where $A$ is unitary and $K$ is the {\em  complex conjugation
  operator}. For the classification according to the anti-unitary
symmetries we can restrict ourselves to the following three different
possibilities: (i) the Hamiltonian does not have any anti-unitary
symmetries, (ii) the Hamiltonian commutes with $AK$ and $(AK)^2 =1$,
and (iii) $[H, AK] = 0$ but $(AK)^2 =-1$.  In the first case, the
matrix elements of the Hamiltonian are complex, in the second case, it
is always possible to find a basis in which the Hamiltonian is real,
and in the third case, it can be shown that it is possible to organize
the matrix elements of the Hamiltonian in quaternion real elements. These
three different possibilities are denoted by the number of degrees of
freedom per matrix element, $\beta =2$, $\beta = 1$, and $\beta = 4$,
respectively.  This {\em  triality} characterizes the Wigner-Dyson
ensembles, the chiral ensembles, and the Altland-Zirnbauer ensembles.
In most cases, the anti-unitary operator is the {\em  time-reversal}
  symmetry operator. For systems without spin, this is just the
complex conjugation operator. For systems with spin, the time reversal
operator can be represented as $i\sigma_2 K$, where $\sigma_2$ is one
of the {\em  Pauli matrices}.

\begin{table}[h*]
\label{tableI}
\centering
\begin{tabular}{cccc}
\hline
  RMT & symmetric space  & $\beta$ & $\alpha$ \\
\hline
 GOE  & AI & 1  & --- \\
 GUE  & A & 2  & --- \\
 GSE  & AII & 4  & --- \\
chGOE  & BDI & 1  & $|\nu|+2N_f$ \\
chGUE  & AIII & 2  & $1+2|\nu|+2N_f$ \\
chGSE  & CII & 4  & $3+4|\nu|+2N_f $\\
AZ-CI  & CI & 1  & 1 \\
AZ-D    & D & 2  & 0 \\
AZ-C    & C & 2  & 2 \\
AZ-DIII    &DIII &  4 & 1\\
\hline
\end{tabular}
\caption{Random matrix ensemble, corresponding symmetric space, and the
values for $\alpha $ and $\beta$.}
\end{table}
\vspace*{0.5cm}

We have introduced ten different random matrix ensembles. Each of
these ensembles can be identified as the {\em tangent space} of one of
the large families of {\em symmetric spaces} as classified by Cartan
(see Table~\ref{tableI}).  The matrices in each of these ten ensembles can be
diagonalized by a unitary transformation, with the unitary matrix
distributed according to the group measure.  For all ensembles, the
{\em Jacobian} for the transformation to eigenvalues as new
integration variables depends only on the eigenvalues.
For an extensive discussion of the calculation of this type of
Jacobian we refer to \cite{HUA}. For the
Wigner-Dyson ensembles, the {\em joint probability distribution} of
the eigenvalues is given by
\be 
P(\{\lambda\}) d \{\lambda\} = {\cal N} |\Delta(\{\lambda\})|^\beta
\prod_k e^{-N\beta \lambda_k^2/4} d\lambda_k,
\label{jointwd}
\ee
where the {\em Vandermonde determinant} is defined by
\be
\Delta(\{\lambda\}) =\prod_{k>l}( \lambda_k -\lambda_l).
\label{vandermonde}
\ee
This factor results in  correlations of eigenvalues that are
characteristic for the random matrix ensembles. For example, one finds
repulsion of eigenvalues at small distances. 

For the remaining ensembles, the eigenvalues occur in pairs $\pm
\lambda_k$.  This results in the distribution
\be 
P(\{\lambda\}) d \{\lambda\} = {\cal N} |\Delta(\{\lambda^2\})|^\beta
\prod_k \lambda_k^\alpha e^{-N\beta \lambda_k^2/4} d\lambda_k.
\ee
The values of $\beta$ and $\alpha$ are given in Table~\ref{tableI}
below. 

Another well-known random matrix ensemble which is not in the above
classification is the {\em Poisson Ensemble} defined as an ensemble of
uncorrelated eigenvalues. Its properties are very different from the
above RMTs where the diagonalization of the matrices leads to strong
correlations between the eigenvalues.

The physical applications of RMT have naturally biased the interest of
researchers to Hermitian matrices (e.g., the Hamiltonian of a quantum
system is a Hermitian operator and should be represented by a
Hermitian matrix).  A variety of
methods, described in this article, have been developed to treat
ensembles of Hermitian matrices. In contrast, {\em non-Hermitian random
matrices} received less attention. Apart from the intrinsic mathematical
interest of such a problem, a number of physically important
applications exist which warrant the study of non-Hermitian random
matrices.

The simplest three classes of non-Hermitian random matrices,
introduced by Ginibre \cite{Ginibre}, are direct generalizations of
the GOE, GUE, and GSE.  They are given by an ensemble of matrices $C$
without any Hermiticity properties and a Gaussian probability
distribution given by
\begin{equation}\label{p(c)}
P(C) DC = {\cal N} e^{-\frac {N\beta}{2} {\rm Tr} C^\dagger C} DC\:,
\end{equation}
where $DC$ is the product of the differentials of the real and imaginary
parts of the matrix elements of $C$.
Such matrices can be diagonalized by a {\em similarity transformation}
with eigenvalues scattered in the complex plane. The probability
distribution is not invariant under this transformation, and therefore
the eigenvalues and the eigenvectors are not distributed
independently.  Similarly to the Hermitian ensembles, the matrix
elements can be chosen real, arbitrary complex, or quaternion real.
 
The case of the arbitrary complex non-Hermitian random matrix ensemble
(\ref{p(c)}) with $\beta=2$ is the simplest.  The joint probability
distribution of eigenvalues
$\{\lambda\}=\{\lambda_1,\ldots,\lambda_N\}$ is given by a formula
similar to (\ref{jointwd}):
\begin{equation}\label{nhjoint}
P(\{\lambda\})d\{\lambda\} = {\cal N} |\Delta(\{\lambda\})|^2 \prod_k
e^{-N|\lambda_k|^2} dx_k dy_k,
\end{equation}
where $x_k={\rm Re\:}\lambda_k$, $y_k={\rm Im\:}\lambda_k$.
In the quaternion-real case, the joint probability distribution can
also be written explicitly.  In the case of real matrices, the joint
probability distribution is not known in closed analytical form.

It is also possible to introduce non-Hermitian ensembles with a chiral
structure, but such ensembles have received very little attention in
the literature and will not be discussed. What has received a great
deal of attention in the literature are {\em non-Hermitian deformations} of
the Hermitian random matrix ensembles. Among others, they enter in the
statistical theory of {\em $S$-matrix fluctuations} \cite{VWZ}, models
of directed quantum chaos \cite{efetov,fyodorov} and in chiral random
matrix models at nonzero chemical potential \cite{Step96}. The latter
class of ensembles is obtained from (\ref{chRMT}) and
(\ref{chRMTprob}) by making the replacement
\be\label{qcd+mu}
\left ( \begin{array}{cc} 0 & C \\ C^\dagger & 0 \end{array} \right )
\rightarrow
\left ( \begin{array}{cc} 0 & C+i\mu \\ C^\dagger +i\mu& 0 \end{array} 
\right ).
\ee
This chRMT is a model for the {\em QCD partition function} at nonzero {\em
  chemical potential} $\mu$ and will be discussed in more detail
below.

Random Matrix Theory is a theory to describe the correlations of the
eigenvalues of a differential operator.  The correlation functions can
be derived from the joint probability distribution.
The simplest object is the {\em spectral density}
\be 
\rho(\lambda) = \sum_k \delta(\lambda-\lambda_k).
\ee
The average spectral density, denoted by
\be
R_1(\lambda) = \langle \rho(\lambda) \rangle,
\ee 
is obtained from the joint probability 
distribution by integration over all eigenvalues except one. The 
{\em connected two-point correlation function} is defined by 
\be
\rho_c(\lambda_1,\lambda_2)
=\langle \rho(\lambda_1) \rho(\lambda_2) \rangle -
\langle \rho(\lambda_1)\rangle \langle \rho(\lambda_2) \rangle .
\ee
In RMT it is customary to subtract the diagonal term from the correlation
function and to introduce the two point correlation function
$R_2(\lambda_1,\lambda_2)$ defined by 
\be
R_2(\lambda_1,\lambda_2) =\langle \rho(\lambda_1) \rho(\lambda_2) \rangle
-\delta(\lambda_1-\lambda_2) \langle \rho(\lambda)\rangle
\ee
and the {\em two-point cluster function}
\be
T_2(\lambda_1,\lambda_2)=  R_1(\lambda_1)R_1(\lambda_2)-
R_2(\lambda_1,\lambda_2).
\ee
In general, the $k-$point correlation function can be expressed in
terms of the joint probability distribution $P_N$ as
\begin{eqnarray}
  \label{Rk}
  R_k(\lambda_1,...,\lambda_k)=
  \frac{N!}{(N-k)!}\int_{-\infty}^\infty
  d\lambda_{k+1}\cdots d\lambda_N P_N(\lambda_1,...,\lambda_N),
\end{eqnarray}
where we have included a combinatorial factor to account for the fact
that spectral correlation functions do not distinguish the ordering of
the eigenvalues.  Similarly, one can define higher order {\em
  connected correlation functions} and cluster functions with all
lower order correlations subtracted out.  For details we refer to
Mehta's book \cite{Meht91}.

Instead of the spectral density one often studies the {\em resolvent} 
defined by
\be
G(z)= \frac 1N {\rm Tr} \frac 1{z-H} = 
\frac 1N \sum_{k=1}^N\frac 1{z-\lambda_k}
\ee
which is related to the spectral density by
\be
\langle\rho(\lambda)\rangle = -
\lim_{\epsilon \rightarrow 0} \frac N{ \pi} 
{\rm Im\:}\langle G(\lambda +i\epsilon)\rangle. 
\label{disc}
\ee

In the analysis of spectra of complex systems and the study of random
matrix theories, it has been found that the average spectral density
is generally not given by the result for the Gaussian random matrix
ensembles which has a semi-circular shape. What is given by RMT are
the correlations of the eigenvalues expressed in units of the average
level spacing.  For this reason one introduces the cluster function
\be
Y_2(r_1,r_2) =\frac {T_2(r_1/R_1(\lambda_1),r_2/R_1(\lambda_2))}
{R_1(\lambda_1) R_1(\lambda_2)}.
\label{Y2}
\ee
In general, correlations of eigenvalues in units of the average level
spacing are called {\em microscopic correlations}. These are the
correlations that can be described by the $N\rightarrow \infty$ limit
of RMT.

The cluster function (\ref{Y2}) has {\em universal properties}. In the
limit $N\rightarrow \infty$, it is invariant with respect to
modifications of the probability distribution of the random matrix
ensemble.  For example, for the GUE and the chGUE it has been shown
that replacing the {\em Gaussian probability distribution} by a distribution
given by the exponent of an arbitrary even polynomial results in the
same microscopic correlation functions \cite{Hack,Damgaard}.

For ensembles in which the eigenvalues occur in pairs $\pm \lambda_k$,
an additional important correlation function with universal properties 
is the {\em microscopic spectral density} \cite{Shur93}  defined by
\be
\rho_s(u) = \lim_{N\rightarrow \infty} \frac 1{\pi\rho(0)} \left\langle
\rho\left(\frac u{\pi\rho(0)}\right)\right\rangle.
\label{rhosu}
\ee 
Related to this observable is the distribution of the smallest
eigenvalue which was shown to be universal as well \cite{Nish98}.  For
this class of ensembles, the point $\lambda = 0$ is a special point.
Therefore, all correlation functions near $\lambda = 0$ have to be
studied separately. However, the microscopic correlations of these
ensembles in the bulk of the spectrum are the same as those of the
Wigner-Dyson ensemble with the same value of $\beta$.

There are two different types of applications of RMT. First, as an
exact theory of spectral correlations of a differential operator.
As an important application we mention the study of universal properties in 
transport phenomena in nuclei \cite{VWZ} and disordered mesoscopic systems. 
In particular, the latter topic has
received a great deal of attention recently (see \cite{Guhr98,Beenreview}). 
This is the original application of RMT.
Second, as a schematic model for a complex system. One famous example
in the second class is the {\em Anderson model} \cite{Anderson} for
{\em Anderson localization}. The properties of this model depend in
a critical way on the spatial dimensionality of the lattice.
Other examples that will be discussed in
more detail below are models for the QCD partition function at
nonzero temperature and nonzero chemical potential.

Random Matrix Theory eigenvalue correlations are not found in all
systems.  Obviously, {\em integrable systems}, for example a {\em
  harmonic oscillator}, have very different spectral properties.
Originally, in the application to nuclear levels, it was believed that
the complexity of the system is the main ingredient for the validity
of RMT.  Much later it was realized that the condition for the
presence of RMT correlations is that the corresponding classical
system is completely {\em chaotic}. This so called 
{\em Bohigas-Giannoni-Schmit
  conjecture} \cite{Bohi84} was first shown convincingly for chaotic
{\em quantum billiards} with two degrees of freedom. By now, this conjecture
has been checked for many different systems, and with some 
well-understood exceptions, it has been found to be correct. However, 
a real proof is still absent, and it cannot be excluded that 
additional conditions may be required for its validity.
In particular, the appearance of collective motion in complex many-body
systems deserves more attention in this respect. 

In general, the average spectral density is not given by RMT.
Therefore, the standard procedure is to unfold the spectrum, i.e., to
rescale the spacing between the eigenvalues according to the local
average eigenvalue density. In practice, this unfolding procedure is
done as follows. Given a sequence of eigenvalues $\{\lambda_k\}$ with
average spectral density $\langle \rho(\lambda)\rangle $, the {\em
  unfolded sequence } is given by
\be
\lambda_k^u = \int_{-\infty}^{\lambda_k} \langle \rho(\lambda) \rangle 
d\lambda.
\ee
The underlying assumption is that the average spectral density and the
eigenvalue correlations factorize. The eigenvalue correlations of the
unfolded eigenvalues can be investigated by means of suitable
statistics. The best known statistics are the {\em nearest neighbor
  spacing distribution} $P(S)$, the {\em number variance}
$\Sigma^2(r)$, and the {\em $\Delta_3$ statistic}.  The number variance
is defined as the variance of the number of eigenvalues in an interval
of length $r$. The $\Delta_3$ statistic is related to the number
variance by
\be
\Delta_3(L) = \frac 2{L^4} \int_0^\infty (L^3 -2L^2r+r^3)\Sigma^2(r) dr.
\ee
In the analysis of spectra it is essential to include only eigenstates
with the same exact quantum numbers. Spectra with different exact
quantum numbers are statistically independent.

The exact analytical expression of the RMT result for the nearest
neighbor spacing distribution is rather complicated. However, it is
well approximated by the {\em Wigner surmise} which is the spacing
distribution for an ensemble of $2\times 2$ matrices. It is given by
\be
P(S) = a_\beta S^\beta e^{-b_\beta S^2},
\ee
where the constants $a_\beta$ and $b_\beta$ can be fixed by the
conditions that $P(S)$ is normalized to unity, and that the average
level spacing is one. The {\em level repulsion} at short distances is
characteristic for interacting systems. For uncorrelated eigenvalues
one finds $P(S) = \exp(-S)$.

Another characteristic feature of RMT spectra is the {\em spectral
stiffness}.  This is expressed by the number variance which,
asymptotically for large $r$, is given by
\be
\Sigma^2(r) \sim \frac 2{\beta \pi^2} \log r.
\ee
This should be contrasted with the result for uncorrelated eigenvalues 
given by $\Sigma^2(r) =r$. 

In the analysis of spectra one often relies on {\em spectral
  ergodicity} defined as the equivalence of spectral averaging and
ensemble averaging. This method cannot be used for the distribution of
the smallest eigenvalues, and one necessarily has to rely on ensemble
averaging.

Before proceeding to the discussion of mathematical methods of Random Matrix
Theory a comment about the notations should be made. 
Different conventions for normalizing the variance
of the probability distribution appear in the literature. This simply
amounts to a rescaling of the eigenvalues.  For example, in the discussion 
of orthogonal polynomials and the Selberg integral below, 
introduction of rescaled eigenvalues such as
$\lambda_k\sqrt{N/2}$ or $\lambda_k\sqrt{N}$ simplifies the expressions.

\section*{Mathematical methods I: Hermitian Matrices}
\label{methods}

\subsection*{Orthogonal polynomials.} 

One of the oldest and perhaps most widely used methods in RMT is based
on {\em orthogonal polynomials}.  A comprehensive presentation of this
method is given in Mehta's book \cite{Meht91}.  Here, we summarize the
most important ingredients, concentrating on the GUE for mathematical
simplicity.

We have seen in the introductory chapter that the spectral correlation
functions can be obtained by integrating the joint probability
distribution.  The mathematical problem consists in performing these
integrations in the limit $N\to\infty$. It is convenient to rescale
$\lambda_k$ and introduce $x_k=\lambda_k\sqrt{N/2}$.  The main
point of the orthogonal-polynomial method is the observation that the
Vandermonde determinant can be rewritten in terms of orthogonal
polynomials $p_n(x)$ by adding to a given row appropriate linear
combinations of other rows,
\begin{equation}
  \Delta(\{x\})=\det\left[(x_j)^{i-1}\right]_{ij}
= {\rm const}\times\det\left[p_{i-1}(x_j)\right]_{ij} \:.
\end{equation}
Including the Gaussian factor in (\ref{jointwd}), this yields
\begin{equation}
  \Delta(\{x\})\prod_{i=1}^Ne^{-x^2/2}={\rm const}\times
  \det\left[\varphi_{i-1}(x_j)\right]_{i,j=1,\dots,N}
\end{equation}
with functions $\varphi_n(x)$ satisfying
\begin{equation}
\int_{-\infty}^\infty dx \:\varphi_m(x)\varphi_n(x) = 
  \int_{-\infty}^\infty dx \: e^{-x^2}p_m(x)p_n(x) =   \delta_{nm}\:.
\end{equation}
In this case, the orthogonal polynomials are essentially the {\em
  Hermite polynomials}, and the $\varphi_n$ are the {\em oscillator
  wave functions},
\begin{equation}
  \varphi_n(x)=\frac{e^{x^2/2}}{\sqrt{2^nn!\sqrt{\pi}}}
  \left(-\frac{d}{dx}\right)^ne^{-x^2}\:.
\end{equation}
The integrals in Eq.~(\ref{Rk}) can now be performed row by row.  The
$k$-point functions are then given by determinants of a two-point
{\em kernel},
\begin{equation}
  R_k(x_1,\dots,x_k)
=\det\left[K_N(x_i,x_j)\right]_{i,j=1,\dots,k}\:.
\label{detstructure}
\end{equation}
The kernel $K_N(x,y)$ is given by
\begin{equation}
  K_N(x,y)=\sum_{n=0}^{N-1}\varphi_n(x)\varphi_n(y)
\end{equation}
which can be evaluated using the {\em Christoffel-Darboux formula}.
In the large-$N$ limit, the spectral density becomes the famous {\em
  Wigner semicircle},
\begin{equation}
  R_1(x)=\lim_{N\to\infty}K_N(x,x)=\frac1\pi\sqrt{2N-x^2}
\end{equation}
if $x^2<2N$ and zero otherwise.  The mean level spacing
$D(x)=1/R_1(x)$ in the bulk of the semicircle thus goes like
$1/\sqrt{N}$.  The $R_k$ are universal if the spacing $|x-y|$ is of
the order of the local mean level spacing, i.e., we require
$|x-y|=rD(x)$ with $r$ of order unity.  In this limit, we obtain
\begin{equation}
  D(x)\lim_{N\to\infty}K_N(x,y)=\frac{\sin(\pi r)}{\pi r}
\end{equation}
which is the famous sine kernel.  The various functions appearing in a
typical RMT-analysis, e.g., $P(s)$, $\Sigma^2(n)$, or $\Delta_3(n)$,
can all be expressed in terms of the $R_k$.

\subsection*{Selberg's integral.} 

In 1944, Selberg computed an integral which turned out to have
significant applications in RMT \cite{Selb44}.  His result reads
\cite{Meht91}
\begin{eqnarray}
  \label{selb}
  I(\alpha,\beta,\gamma,n) 
  &=&\int_0^1dx_1\cdots\int_0^1dx_n |\Delta(x)|^{2\gamma}
  \prod_{j=1}^nx_j^{\alpha-1}(1-x_j)^{\beta-1} \nonumber\\
  &=& \prod_{j=0}^{n-1}\frac{\Gamma(1+\gamma+j\gamma)
    \Gamma(\alpha+j\gamma)\Gamma(\beta+j\gamma)}{\Gamma(1+\gamma)
    \Gamma(\alpha+\beta+(n+j-1)\gamma)}\:,
\end{eqnarray}
where $\Delta(x)$ is the Vandermonde determinant, $n$ is an integer,
and $\alpha$, $\beta$, and $\gamma$ are complex numbers
\mbox{satisfying} ${\rm Re\:}\alpha >0$, ${\rm Re\:}\beta >0$,
\mbox{${\rm Re\:}\gamma > -\min\{1/n,{\rm Re\:}\alpha/(n-1)$,} ${\rm
  Re\:}\beta/(n-1)\}$.  Choosing the parameters in Eq.~(\ref{selb})
appropriately, one can derive special forms of Selberg's integral
related to specific orthogonal polynomials \cite{Meht91}.  For
example, choosing $x_i=y_i/L$, $\alpha=\beta=aL^2+1$, and taking the
limit $L\to\infty$, one obtains the integrals of the joint probability
density function of the GUE which are related to Hermite polynomials.
Selberg's integral is also very useful in the derivation of {\em
  spectral sum rules} \cite{Verb94d}.

Aomoto derived the following generalization of Selberg's integral
\cite{Aomo87}, 
\begin{eqnarray}
  \label{aomo}
  \int_0^1dx_1&\cdots&\int_0^1dx_n \:x_1\cdots x_m 
    |\Delta(x)|^{2\gamma}\prod_{j=1}^nx_j^{\alpha-1}
    (1-x_j)^{\beta-1} \nonumber\\
  &=&\prod_{j=1}^m\frac{\alpha+(n-j)\gamma}{\alpha+\beta+(2n-j-1)\gamma}
    \:I(\alpha,\beta,\gamma,n)\:,
\end{eqnarray}
where $1\le m\le n$.  A further extension of Selberg's integral was
considered by Kaneko \cite{Kane93} who related it to a system of
partial differential equations whose solution can be given in terms of
{\em Jack polynomials}.

\subsection*{Supersymmetric method}

The supersymmetric method has been applied successfully to 
problems where the orthogonal polynomial method has failed
\cite{Efetov,Efetovbook,VWZ}.
It relies on the observation that the average {\em resolvent} 
can be written as
\be
\langle{ G(z)}\rangle = \frac 1N\left\langle {\rm Tr }\frac 1{z-H} 
\right\rangle 
= \frac 1N\left . \frac {\partial}{\partial J}\right |_{J=0}
Z(J),
\ee
where the {\em generating function} is defined by
\be
Z(J) = \int DH P(H)\frac {\det (z-H+J)}{\det(z-H)},
\ee
and the integral is over the probability distribution of one of the
random matrix ensembles defined in the introductory chapter.  The
determinant can be expressed in terms of Gaussian integrals,
\be
\frac {\det (z-H+J)}{\det(z-H)} &=& \int d\{ \psi \}
\exp\biggl( \sum_{kl}[i\phi_k^* (z-H)_{kl} \phi_l
  +i\chi_k^* (z+J-H)_{kl} \chi_l]\biggr)\:,
\ee
where the measure is defined by
\be
d\{\psi\} = \prod_{j=1}^N {d\phi_j d\phi^*_j d\chi_j
d\chi^*_j\over2\pi}\ .
\ee
For convergence the imaginary part of $z$ has to be positive. The
integrations over the real and imaginary parts of $\phi_i$ range over
the real axis (the usual commuting, or {\em bosonic} variables), 
whereas $\chi_i$ and $\chi^*_i$ are {\em Grassmann
  variables} (i.e., anticommuting, or {\em fermionic} variables) 
with integration defined according to the convention that
\be
\int d\chi = 0 \quad {\rm and} \quad \int \chi d\chi = 1\ . 
\label{grassint}
\ee
With this normalization, $Z(0) = 1$.

For simplicity, we consider the GUE [$\beta = 2$ in (\ref{probwd})]
which mathematically is the simplest ensemble.  The Gaussian integrals
over $H$ can be performed trivially, resulting in the the generating
function 
\be 
Z(J)&=& \int d \{\psi\} \exp\Biggl[ -\frac 1{2N} {\rm Trg}  \left (
\begin{array}{cc} \sum_j \phi_j^* \phi_j & \sum_j \chi_j^* \phi_j\\
\sum_j \chi_j \phi_j^*  & \sum_j \chi_j^* \chi_j \end{array} \right )^2
 \nonumber \\ &&\hspace*{25mm}
+ i\sum_j (\phi_j^* z \phi_j + \chi_j^*(z+J)\chi_j) \Biggr],
\ee 
where the sums over $j$ run from $1$ to $N$.  The symbol Trg denotes
the {\em graded trace} (or {\em supertrace}) 
defined as the difference of the trace of
the boson-boson block (upper left) and the trace of the
fermion-fermion (lower right) block. For example, in terms
of the $2\times 2$ matrix (\ref{sigma}), 
${\rm Trg} \sigma =  \sigma_{BB} -  i\sigma_{FF}$.  
The quartic terms in $\phi$ and
$\chi$ can be expressed as Gaussian integrals by means of a {\em
  Hubbard-Stratonovitch transformation}. This results in
\be
Z(J)&=& \int d \{\psi\} d\sigma
\exp\Biggl[ -\frac N2{\rm Trg\:} \sigma^2 + 
i\sum_j\left (\begin{array}{c} \phi_j^* \\ 
\chi_j^* \end{array} \right )
(\sigma + \zeta)
 \left (\begin{array}{c} \phi_j \\ \chi_j \end{array} \right )\Biggr],
\ee 
where 
\be
\sigma =
\left (\begin{array}{cc} \sigma_{BB} & \sigma_{BF} \\ \sigma_{FB} & 
i\sigma_{FF} \end{array} \right )
\label{sigma}
\ee
and
\be
\zeta =
\left (\begin{array}{cc} z & 0 \\ 0 & 
z + J \end{array} \right ).
\ee
The variables $\sigma_{BB}$ and $\sigma_{FF}$ are commuting (bosonic) 
variables that range over the full real axis. Both $\sigma_{BF}$ and
$\sigma_{FB}$ are Grassmann (fermionic) variables.

The integrals over the $\phi$ and the $\chi$ variables are now
Gaussian and can be performed trivially. This results in the
$\sigma$-model
\be
Z(J)= \int d\sigma
\exp\left[ -\frac N2{\rm Trg} \sigma^2 + N{\rm Trg } \log (\sigma +
\zeta)\right].
\ee 
By shifting the integration variables according to $\sigma \rightarrow
\sigma -\zeta$ and carrying out the differentiation with respect to
$J$ one easily finds that
\be
\langle{G(z)}\rangle = \langle z-i\sigma_{FF} \rangle.
\ee
In the large $N$ limit, the expectation value of $\sigma_{FF} $ 
follows from a {\em saddle-point }
analysis.
The saddle point equation for $\sigma_{FF}$ is given by
\be
\sigma_{FF} + i z = 1/\sigma_{FF} 
\ee
resulting in the resolvent 
\be
\langle G(z)\rangle = \frac z2 - \frac i2 \sqrt{4 -z^2}.
\label{result}
\ee
Using the relation (\ref{disc}) we find that the average spectral
density is a semi-circle.

The supersymmetric method can also be used to calculate spectral
correlation functions.  They follow from the average of the {\em
  advanced} and the {\em retarded} resolvent. In that case we do not
have a saddle-point but rather a {\em saddle-point manifold} related
to the {\em hyperbolic symmetry} of the retarded and advanced parts of
the generating function. The supersymmetric method not only provides us
with  alternative derivations of known results. As an example we mention
that the analytical result for $S$-matrix fluctuations at different energies
was first derived by means of this method \cite{VWZ}.

Alternatively, it is possible to perform the $\sigma$ integrations by
a supersymmetric version of the {\em Itzykson-Zuber integral}
\cite{Guhr91} rather than a saddle-point approximation.  The final
result is an exact expression for the kernel of the correlation
functions. The advantage of this method is that it exploits the
determinantal structure of the correlation functions [see
(\ref{detstructure})] and all correlations functions are obtained at
the same time. Moreover, the results are exact at finite $N$.

\subsection*{Replica trick}

The {\em replica trick}, which was first introduced in the
theory of {\em spin glasses} \cite{Edwards}, is based on the
observation that 
\be
\left\langle {\rm Tr }\frac 1{z-H} \right\rangle = \lim_{n\rightarrow 0}
\frac 1n\left . \frac {\partial}{\partial J}\right |_{J=0}
Z_n(J),
\ee
where the generating function is defined by
\be
Z_n(J) = \int DH P(H) {{\det}^n (z-H+J)}.
\ee
The determinant can then be expressed as a Grassmann integral where
the $\chi$-variables now have an additional flavor index,
\begin{equation}\label{replicas}
{\det}^n (z-H+J) = \int d\{\chi\} \exp \sum_{ij} \chi^{f*}_i 
(z-H+J)_{ij} \chi^f_j.
\end{equation}
The sum over $f$ ranges from $1$ to $n$,
and the measure is defined by
\be
d\{\chi\} = \prod_{f=1}^n \prod_{i=1}^N d\chi_i^f d\chi^{f*}_i.
\ee
After averaging over the matrix elements of $H$ and a
Hubbard-Stratonovitch transformation one can again proceed to the
$\sigma$ variables. In this case, we only have a $\sigma_{FF}$ block
which is now an $n\times n$ matrix. The average resolvent then follows
by making a saddle point approximation and taking the replica limit
with the same final result as given in (\ref{result}).

Because the replica trick relies on an {\em analytical continuation}
in $n$, it is not guaranteed to work. Several explicit examples for
its failure have been constructed \cite{VZ,Step96}.  In general, it
cannot be used to obtain nonperturbative results for eigenvalue
correlations on the microscopic scale which decreases as $1/N$ in the
limit $N\rightarrow \infty$.

\subsection*{Resolvent Expansion Methods}

The Gaussian averages can also be performed easily by expanding the
resolvent in a geometric series in $1/z$,
\be
G(z) = \frac 1z + \frac 1N{\rm Tr}\frac 1z H \frac 1z +  
\frac 1N{\rm Tr}\frac 1z H \frac 1z H \frac 1z + \cdots.
\ee
The Gaussian integral over the probability distribution of the
matrix elements is given by the sum over all {\em pairwise
contractions}. For the GUE, a contraction is defined as
\be
\langle{ H_{ij} H^\dagger_{kl}}\rangle = \frac 1N \delta_{il} \delta_{jk} .
\ee
To leading order in $1/N$, the 
contributions are given by the nested contractions. One easily derives that
the average resolvent satisfies the equation 
\be
\langle G(z)\rangle = \frac 1z(1 + \langle G(z) \rangle^2 ),
\ee
again resulting in the same expression for the average resolvent.

This method is only valid if the geometric series is convergent. For
this reason, the final result is only valid for the region that 
can be reached from large values of $z$ by analytical continuation. For
non-Hermitian matrices this leads to the failure of this method, and instead
one has to rely on the so-called {\em Hermitization}.

As is the case with the replica trick, this method does not work to
obtain nonperturbative results for microscopic spectral correlations. 
This method has been used widely in the literature. As one of the earlier
references we mention the application to the statistical theory of
nuclear reactions \cite{Agassi}

\subsection*{Dyson Gas}

The formula (\ref{jointwd}) suggests a very powerful analogy between the
{\em Wigner-Dyson random matrix ensembles} and the statistical properties
of a gas of charged particles restricted to move in one dimension, 
the {\em Dyson gas}~\cite{Dyson}. 

Let $\lambda_k$ be a coordinate of a classical 
particle which moves in the potential
$V_1(\lambda_k)=N\lambda_k^2/4$. 
Furthermore, let two such particles repel
each other so that the potential of the {\em pairwise interaction}
is $V_2(\lambda_k,\lambda_l)=-\ln|\lambda_k-\lambda_l|$. 
If one considers a gas of $N$
such particles in thermal equilibrium at temperature $T$,
then the probability distribution for the coordinates
of the particles $\lambda=\{\lambda_1,\ldots,\lambda_N\}$ will be 
proportional to $\exp(-V(\lambda)/T) \prod_k d\lambda_k$, where
the potential energy $V$ is given by
\begin{equation}
V(\lambda)=\sum_{k<l} V_2(\lambda_k,\lambda_l) + \sum_i V_1(\lambda_i).
\end{equation}
If the temperature $T$ of the gas is chosen to be equal to $1/\beta$,
the probability distribution of the coordinates of the particles
becomes identical to the probability distribution (\ref{jointwd}) of the
eigenvalues.

This analogy allows one to apply methods of {\em statistical mechanics} to
calculate distributions and correlations of the eigenvalues
\cite{Dyson}.  It also helps to grasp certain aspects of {\em universality}
in the statistical properties of the eigenvalues.  In particular, it
is understandable that the correlations in the relative positions of
particles are determined by the interactions between them, i.e., by
$V_2$, and are generally insensitive to the form of the
single-particle potential, $V_1$. On the other hand, the overall
density will depend on the form of the potential $V_1$.

The logarithmic potential $V_2$ is the {\em Coulomb potential} in 
two-dimensional space (i.e., it satisfies the two-dimensional {\em Laplace
equation} $\Delta V_2 = 0$). Therefore, the Dyson gas can be viewed as
a two-dimensional Coulomb gas, with the kinematic restriction that
the particles move along a straight line only. This restriction is
absent in the case of non-Hermitian matrices.

\section*{Mathematical methods II: non-Hermitian matrices}

The eigenvalues of {\em non-Hermitian matrices} are not constrained to lie
on the real axis. Instead, they occupy the two-dimensional {\em complex
plane}.  This fact requires non-trivial modifications of some of the
methods developed for Hermitian matrices.  Surprisingly or not, the
required formalism is sometimes simpler, and sheds more light on the
properties of Hermitian random matrices.

\subsection*{Orthogonal polynomials}

The method of orthogonal polynomials can also be applied to treat
non-Hermitian random matrices. The simplest example is the {\em Ginibre
ensemble} of arbitrary complex matrices (\ref{p(c)}) with
$\beta=2$~\cite{Ginibre}. It is convenient to
rescale $\lambda_k$ and introduce $w_k=\lambda_k\sqrt N$.
The orthogonal
polynomials with respect to the weight given by $\exp(-|w|^2)$
are simply the monomials $w^n$. Indeed,
\begin{equation}
\int dudv e^{-|w|^2} w^n (w^*)^m = \pi n! \delta_{mn},
\end{equation}
where $w = u+iv$.
The orthonormal functions, $\int dudv\phi_n(w)\phi_m(w^*)=\delta_{nm}$,
are, therefore,
\begin{equation}
\phi_n(w) = {1\over \sqrt{\pi n!}} e^{-|w|^2/2} w^n.
\end{equation}
Following the same steps as in the case of the Hermitian GUE one
obtains all correlation functions in the form of the determinant
\begin{equation}
R_k(w_1,\ldots,w_k) = \det [K_N(w_i, w_j)]_{i,j=1,\ldots,k},
\end{equation}
with a kernel $K_N$ given by
\begin{equation}
K_N(w_1,w_2) = \sum_{n=0}^{N-1} \phi_n(w_1) \phi_m(w_2^*).
\end{equation}
By a careful analysis of the large-$N$ limit of the kernel one finds that
$R_1(w)$ is $1/\pi$ inside the complex disk  $|w| < \sqrt N$ and vanishes
outside this domain.

\subsection*{Coulomb gas}

The probability distribution (\ref{nhjoint}) is the same as for a 
{\em Coulomb gas} in two dimensions placed in the {\em harmonic potential}
$V_1=N|z|^2/2\equiv N(x^2+y^2)/2$ at a temperature $1/\beta=1/2$.
Unlike in the Hermitian case, the particles of the gas are now allowed
to move in both dimensions.

The analogy with the Coulomb gas can be used to calculate the density
of eigenvalues of the ensemble of complex non-Hermitian matrices
[(\ref{p(c)}) with $\beta=2$] in the limit $N\to\infty$. In this
limit, the typical energy per particle, $O(N)$, is infinitely larger
than the temperature, 1/2. Therefore, the system is assuming an
equilibrium configuration with the minimal energy, as it would at zero
temperature. Each particle is subject to a linear force
$-dV_1/d|z|=-N|z|$ directed to the origin, $z=0$.  This force has to
be balanced by the Coulomb forces created by the distribution of all
other particles. Thus, the electric field created by this distribution
must be directed along the radius and be equal to $|\vec E|=N|z|$.
The {\em Gauss law}, $\vec\nabla \vec E = 2\pi\rho$, tells us that such a
field is created by charges distributed uniformly (with density
$\rho=N/\pi$) inside a circle around $z=0$, known as the {\em Ginibre
  circle}.  The radius of this circle, $R$, is fixed by the total
number of the particles, $\pi R^2 \rho=N$, so that $R=1$.

\subsection*{Electrostatic analogy and analyticity of resolvent}

In general, the mapping of the random matrix model onto the Coulomb
gas is not possible, because the {\em pairwise interaction} is not
always given simply by the logarithm of the distance between the particles.
However, a more generic electrostatic analogy exists, relating
the {\em two-dimensional density of eigenvalues} $\rho$,
\begin{equation}
\rho(x,y) = \sum_k \delta(x-x_k) \delta(y-y_k),
\end{equation}
where $x_k$ and $y_k$ are real and imaginary parts of $\lambda_k$,
and the {\em resolvent} $G$,
\begin{equation}
G(x,y) = {1\over N} {\rm Tr} {1\over z-C} 
= {1\over N} \sum_k {1\over z-\lambda_k}.
\end{equation}
Since the electric field created by a point charge in two dimensions
is inversely proportional to the distance from the charge, one can see
that the two-component field $(N{\rm Re\:}G, -N{\rm Im\:}G)$ coincides
with the electric field $\vec E$, created by the charges located at
the points $\{\lambda_1,\ldots,\lambda_N\}$ in the {\em complex plane}.

The {\em Gauss law}, relating the density of charges and the resulting
electric field, $\vec \nabla \vec E = 2\pi\rho$, gives the following
relation between the density of the eigenvalues and the resolvent:
\begin{equation}\label{estatic}
\rho = {N\over2\pi} \left\{{\partial{\rm Re\:}G\over\partial x} 
- {\partial{\rm Im\:}G\over\partial y}\right\}\:.
\end{equation}
This relation is the basis of methods for the calculation of  the
average density of the eigenvalues, $\langle \rho \rangle$. The r.h.s
of this equation vanishes if $G$ obeys the {\em Cauchy-Riemann conditions},
i.e., if it is an {\em analytic function} of the complex variable $z=x+iy$.
Conversely, $\rho$ describes the location and the amount of
non-analyticity in $G$.

In the case of Hermitian matrices, $C=H$, the eigenvalues lie on a
line (real axis) and, after {\em ensemble averaging}, they fill a continuous
interval. This means that the average resolvent $\langle G(x,y)
\rangle$ has a {\em cut} along this interval on the real axis.  The
discontinuity along this cut is related to the {\em linear density} of
the eigenvalues by (\ref{disc}).  In the case of a non-Hermitian
matrix $C$, the eigenvalues may and, in general, do fill
two-dimensional regions. In this case, the function $G$ is not
analytic in such regions.

This is best illustrated by the ensemble of arbitrary complex matrices
$C$ (\ref{p(c)}). In the $N\to\infty$ limit, the resolvent is given by
\begin{equation}\label{gcircle}
G(x,y) = \left\{
\begin{array}{ll}
z^*, & |z|<1\\
1/z, & |z|>1\\
\end{array}
\right. .
\end{equation}
One observes that $G$ is non-analytic inside the {\em Ginibre
circle}~\cite{Ginibre}.

\subsection*{Replica trick}

The generalization of the {\em replica trick} to the case of non-Hermitian
matrices is based on the relation
\begin{equation}\label{nhreplica}
\left\langle {\rm Tr} {1\over z-C} \right\rangle 
= \lim_{n\to0} \frac 1n {\partial\over\partial z} \ln Z_{n}(z),
\end{equation}
where now
\begin{equation}\label{|det|}
Z_{n}(z) = \int DC P(C) |{\det} (z-C)|^n. 
\end{equation}
The absolute value of the determinant can be also written as
${\det}^{n/2} (z-C) {\det}^{n/2} (z^*-C^\dagger)$.
Following (\ref{replicas}), one introduces $n/2$ Grassmann variables
$\chi_i$ to represent ${\det}^{n/2} (z-C)$ and another $n/2$
to represent ${\det}^{n/2} (z^*-C^\dagger)$. If the measure
$P(C)$ is Gaussian, the integral over $C$ can now be performed
resulting in terms quartic in the Grassmann variables. These can be
rewritten with the help of an auxiliary $n\times n$ variable $\sigma$ 
as bilinears in $\chi$, after which the $\chi$ integration can be
done. The resulting integral, in the limit $N\to\infty$,
is given by the {\em saddle point} (maximum) of its integrand. In the
case of the Ginibre ensembles one arrives at the following expression:
\begin{equation}
\ln Z_n(z) = nN \max_\sigma 
\left[ - |\sigma|^2 + \ln(|z|^2 + |\sigma|^2) \right].
\end{equation}
There are two possible maxima, $\sigma=0$ and $|\sigma|^2=1-|z|^2$,
which give two branches for $\ln Z_n$, $\ln Z_n/(nN)=\ln|z|^2$ and 
$\ln Z_n/(nN)=|z|^2-1$. The former dominates when $|z|>1$
and the latter when $|z|<1$. Using (\ref{nhreplica}) one obtains the
average resolvent given by (\ref{gcircle}).

It is important that the absolute value of the determinant is taken
in (\ref{|det|}). Without taking the absolute value one would obtain the 
incorrect result
$G=1/z$ everywhere in the complex plane.

\subsection*{Hermitization}

The method of {\em Hermitization}, as well as the replica trick 
(\ref{nhreplica},\ref{|det|}), is based on the observation that 
$\Delta \ln |z|^2 = 4\pi\delta(x)\delta(y)$,
where $\Delta$ is the Laplacian in the coordinates $x$ and $y$.
One can, therefore, write for the eigenvalue density $\rho(x,y)$
\begin{equation}\label{hermtn}
4\pi\langle\rho(x,y)\rangle = 
\Delta \langle \ln\det(z-C)(z^*-C^\dagger) \rangle.
\end{equation}
The determinant on the r.h.s. can be written as the
determinant of a matrix (up to a sign)
\begin{equation}
H(z,z^*) =
\left(
\begin{array}{cc}
0               &       z-C     \\
z^* - C^\dagger &       0       \\
\end{array}
\right).
\end{equation}
This matrix is Hermitian, and one can apply methods of Hermitian
RMT (e.g., the supersymmetric method or the replica trick)
to determine its resolvent 
$G(\eta)$. Integrating over $\eta$ one obtains
the quantity
\begin{equation}
\langle \ln\det(\eta-H)\rangle,
\end{equation}
which in the limit $\eta\to0$ reduces to the expression on the
r.h.s. of (\ref{hermtn}) \cite{Girko,efetov,sompolinsky,zee}.

\section*{Applications and advanced topics}
\label{advanced}

In this section, we briefly review a variety of different subfields of
physics where RMT has been applied successfully.  Most of the examples
can be found in the comprehensive presentation of Ref.~\cite{Guhr98}
which also contains a wealth of useful references.

\subsection*{Nuclear level spacings.}

\begin{figure}[ht]
\unitlength 1cm
\begin{picture}(14.0,6)
  \centerline{\includegraphics[width=110mm]{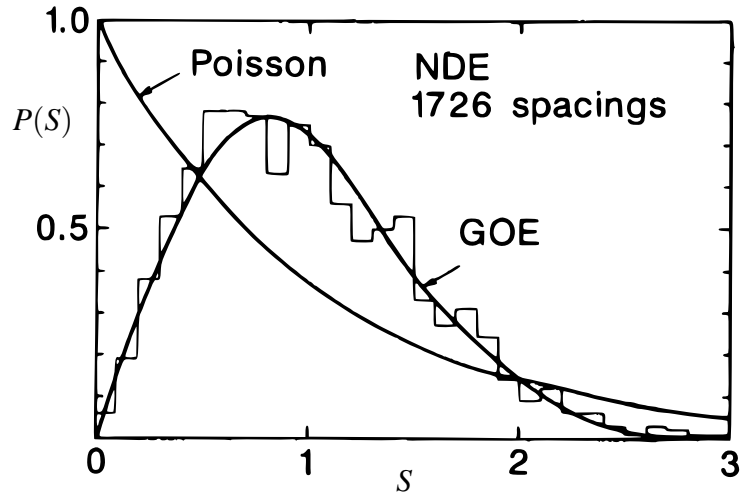} }
\caption{ The histogram represents the nearest-neighbor spacing
    distribution of the ``Nuclear Data Ensemble'' (NDE).  The curve
    labeled GOE is the random-matrix prediction, and the Poisson
    distribution, representing uncorrelated eigenvalues, is shown for
    comparison.  Taken from Ref.~\protect\cite{Bohi83} with kind permission
    from Kluwer Academic Publishers.}%
\label{Fig1}
\put(-7.4,1){{\sf\bf $S$}}
\put(-12.5,5.8){{\sf{\bf $P(S)$}}}
\end{picture}
\end{figure}

Historically, the first application of
RMT in physics arose in the study of {\em nuclear energy levels}.  The
problem of computing highly excited energy levels of large nuclei is
so complicated that it is impossible to make detailed predictions
based on microscopic models.  Therefore, as discussed in the
introduction, it is interesting to ask whether the statistical
fluctuations of the nuclear energy levels are universal and described
by the predictions from RMT.  The nuclear Hamiltonian is {\em time-reversal
invariant} so that the data should be compared with GOE results.
Figure~\ref{Fig1} shows the nearest neighbor spacing distribution of
nuclear energy levels of the same spin and parity, averaged over 1726
spacings from 32 different nuclei \cite{Bohi83}.  Clearly, the data
are described by RMT, indicating that the energy levels are strongly
correlated.  The parameter-free agreement seen in the figure gave
strong support to the ideas underlying RMT.

\subsection*{Hydrogen atom in a magnetic field.} 

\begin{figure}[ht!]
  \centerline{\scalebox{0.5}{\includegraphics[width=160mm]{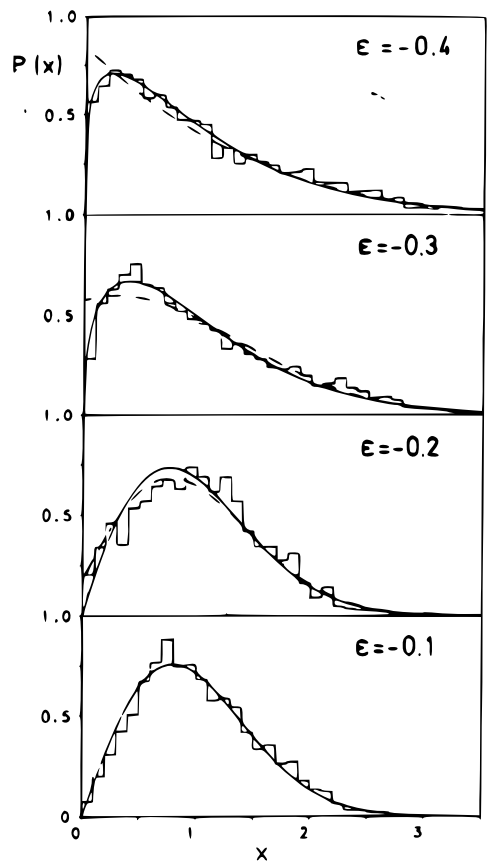}}}
\caption{ Nearest neighbor spacing distribution of the energy levels
    of the hydrogen atom in a magnetic field (histograms).  The solid
    line in the bottom plot is the RMT prediction for the GOE, all
    other lines are fits.  As a function of the scaled variable
    $\varepsilon$, which increases from top to bottom, a transition
    from Poisson [$P(x)=\exp(-x)$] to RMT behavior is observed.  Taken
    from Ref.~\protect\cite{Frie89} with kind permission from 
    Elsevier Science.}
\label{Fig2}
\end{figure}

The Hamiltonian of this system is given by
\begin{equation}
  H=\frac{{\bf p}^2}{2m}-\frac{e^2}{r}-\omega L_z
  +\frac12\omega^2(x^2+y^2)\:, 
\end{equation}
where $m$ is the reduced mass, $e$ is the unit charge,
$r=(x^2+y^2+z^2)^{1/2}$ is the separation of proton and electron,
$\omega=eB/(2mc)$ is the {\em Larmor frequency}, $B$ is a constant magnetic
field in the $z$-direction, and $L_z$ is the third component of the
angular momentum.  At $B=0$, the system is integrable.  This property
is lost when the magnetic field is turned on, and large parts of the
classical phase space become chaotic.  For an efficient numerical
computation of the eigenvalues, it was important to realize that the
Hamiltonian has a scaling property which simplifies the calculations
considerably: The spectrum depends only on the combination
$\varepsilon=\gamma^{-2/3}E$, where $\gamma$ is a dimensionless
variable proportional to $B$ and $E$ is the energy of the system
\cite{Wint87}.  This variable increases if the magnetic field is
increased and/or the ionization threshold is approached from below.
Thus, as a function of $\varepsilon$, one should observe a transition
from Poisson to RMT behavior in the spectral correlations.  The
numerical results are in agreement with experimental data and clearly
show a Poisson to RMT transition, see Fig.~\ref{Fig2}.

\subsection*{Billiards and quantum chaos.} 

These are the prototypical systems
used in the study of {\em quantum chaos}.  A billiard is a dynamical system
consisting of a point particle which can move around freely in a
bounded region of space.  For simplicity, we assume that the space is
two-dimensional.  In a {\em classical billiard}, the particle is reflected
elastically from the boundaries corresponding to a potential that
is zero inside the boundary and infinite outside the boundary. In a 
{\em quantum billiard} this results in a free particle 
{\em Schr\"odinger equation}
with wave functions that vanish on and outside this boundary.
Depending on the shape of the boundary, the classical motion of the
particle can be regular or chaotic (or mixed).  Examples of
classically regular billiards are the rectangle and the ellipse.
Important classically chaotic billiards are the stadium billiard
(i.e., two semicircles at opposite sides of an open rectangle) and the
{\em Sinai billiard} (i.e., the region outside a circle but inside a
concentric square surrounding the circle).  According to the
conjecture by Bohigas, Giannoni, and Schmit \cite{Bohi84}, the level
correlations of a quantum billiard whose classical counterpart is
chaotic should be given by RMT, whereas the eigenvalues of a quantum
billiard whose classical analog is regular should be uncorrelated and,
thus, described by a Poisson distribution.  This conjecture was
investigated --- numerically, semiclassically, or using periodic orbit
theory --- in a number of works and confirmed in almost all cases
\cite{Gutz90}.  One can also vary the shape of a billiard as a
function of some parameter, thus interpolating between a classically
regular and a classically chaotic billiard.  As a
function of the parameter, one then observes a transition from Poisson
to RMT behavior in the level correlations of the corresponding quantum
billiard.

\subsection*{Quantum dots.}  

{\em Semiconducting microstructures} can be fabricated such that the
electrons are confined to a two-dimensional area.  If this region is
coupled to external leads, we speak of {\em a quantum dot}.  Such systems
have many interesting properties.  If the elastic mean free path of
the electrons (which at very low temperatures is
${\scriptscriptstyle\stackrel>\sim}10\,\mu$m) is larger than the
linear dimensions ($\sim 1\,\mu$m) of the quantum dot, and if the
Coulomb interaction is neglected, the electrons can move around freely
inside the boundary, and the quantum dot can be thought of as a
realization of a {\em quantum billiard}.  Depending on the shape of the
quantum dot, certain observables, e.g., the fluctuations of the
conductance as a function of an external magnetic field, show a
qualitatively different behavior.  If the shape is classically chaotic
(e.g., a stadium), the experimental results agree with predictions
from RMT as expected, in contrast to data obtained with quantum dots
of regular shape where the fluctuations are not universal
\cite{Marc92}. For a recent review of quantum dots and universal
conductance fluctuations to be discussed in the following section we
refer to Ref.~\cite{Beenreview}.

\subsection*{Universal conductance fluctuations.}  

A {\em mesoscopic system} in
condensed matter physics is a system whose linear size is larger than
the {\em elastic mean free path} of the electrons but smaller than the 
{\em phase
coherence length}, which is essentially the {\em inelastic mean free path}.
A typical size is of the order of 1 $\mu$m.  The {\em conductance}, $g$, of
mesoscopic samples is closely related to their spectral properties.
Using a scaling block picture, Thouless found that in the {\em diffusive
regime}, $g=E_C/\Delta$, where $E_C/\hbar$ is the inverse {\em diffusion
time} of an electron through the sample and $\Delta$ is the mean level
spacing \cite{Thou74}.  This can be rewritten as $g=\langle
N(E_C)\rangle$, where $\langle N(E)\rangle$ is the mean level number
in an energy interval $E$.  Thus the variance, $\langle\delta
g^2\rangle$, of the conductance is linked to the number variance,
$\Sigma^2$, of the energy levels.

In experiments at very low temperatures where the conductance of
mesoscopic wires was measured as a function of an external magnetic
field, people have observed fluctuations in $g$ of the order of
$e^2/h$, independent of the details of the system (shape, material,
etc.).  These are the so-called {\em universal conductance
  fluctuations} \cite{Wash86}.  This phenomenon can be understood
qualitatively by estimating the number fluctuations of the electron
levels using RMT results.  However, the magnitude of the effect is
much larger than expected, due to complicated quantum interference
effects.  While a truly quantitative analysis requires {\em linear
  response theory} (the {\em Kubo formula}) or the {\em multichannel
  Landauer formula}, both the magnitude of the fluctuations as well
their universality can be obtained in a simpler approach using the
{\em transfer matrix method}.  Here, the assumption (although not
quantitatively correct) is that certain parameters of the transfer
matrix have the same long-range stiffness as in RMT spectra.

\subsection*{Anderson Localization}

{\em Anderson localization} is the phenomenon that a good {\em conductor}
becomes an {\em insulator} when the disorder becomes sufficiently
strong. Instead of a description of the electron wave functions by
{\em Bloch waves}, the wave function of an electron becomes {\em
  localized} and decays exponentially, i.e.,
\be
\psi(r) \sim e^{-  r/{L_c}}.
\ee 
The length scale $L_c$ is known as the {\em localization length}.
This phenomenon was first described in the Anderson model
\cite{Anderson} which is a {\em hopping model} with a {\em random potential}
on each lattice point. The dimensionality of the lattice plays an
important role.  It has been shown that in one dimension all states
are localized. The critical dimension is two, whereas for $d= 3$ we
have a {\em delocalization transition } at an energy $E_L$. All states
below $E_L$ are {\em localized} whereas all states above $E_L$ are
{\em extended}, i.e., with a wave function that scales with the
size of the system.

The eigenvalues of the localized states are not correlated, and their
correlations are described by the Poisson distribution. In the
extended domain the situation is more complicated. An important energy
scale is the {\em Thouless energy} \cite{Thou74} which is related to
the diffusion time of an electron through the sample. With the latter
given by $L^2/D$ (the {\em diffusion constant} is denoted by $D$) this
results in a Thouless energy given by
\be
E_c = \frac{\hbar D}{L^2}.
\ee
Correlations on an energy scale below the Thouless energy are given by
Random Matrix Theory, whereas on higher energy scales the eigenvalues
show weaker correlations.

\subsection*{Other wave equations.}  

So far, we have implicitly considered
quantum systems which are governed by the {\em Schr\"odinger equation}.  It
is an interesting question to ask if the eigenmodes of systems obeying
{\em classical wave equations} display the same spectral fluctuation
properties as predicted by RMT.  Classical wave equations arise, e.g.,
in the study of {\em microwave cavities} or in {\em elastomechanics} and
{\em acoustics}.

In three-dimensional microwave cavities, the electric and magnetic
fields are determined by the {\em Helmholtz equation}, $(\Delta^2+{\bf
  k}^2){\bf A}({\bf r})=0$, where ${\bf A}={\bf E}$ or ${\bf B}$.  It
was found experimentally that the spacing of the eigenmodes of the
system is of RMT type if the cavity has an irregular shape
\cite{Deus95}.  If the cavity has some regular features, the spacing
distribution interpolates between RMT and Poisson behavior
\cite{Alt96}.

Elastomechanical eigenmodes have been studied both for aluminum and for
quartz blocks.  Here, there are two separate Helmholtz equations for
the longitudinal (pressure) and transverse (shear) waves,
respectively, making the problem even more different from the
Schr\"odinger equation.  Several hundred up to about 1500 eigenmodes
could be measured experimentally.  A rectangular block has a number of
global symmetries, and the measured spectrum is a superposition of
subspectra belonging to different symmetries.  In such a situation,
the spacing distribution of the eigenmodes is expected to be of
Poisson type, and this was indeed observed experimentally.  The
symmetry can be broken by cutting off corners of the block, and the
resulting shape is essentially a three-dimensional {\em Sinai billiard}.
Depending on how much material was removed from the corners, a Poisson
to RMT transition was observed in the spacing distribution of the
eigenmodes \cite{Elle95}.

Thus, we conclude that RMT governs not only the eigenvalue
correlations of the Schr\"odinger equation but also those of rather
different wave equations.

\subsection*{Zeros of the Riemann zeta function.} 

This is an example from number theory which, at first sight, is not
related to the theory of dynamical systems.  The {\em Riemann zeta
  function} is defined by $\zeta(z)=\sum_{k=1}^\infty k^{-z}$ for
${\rm Re\:}z>1$.  Its nontrivial zeros $z_n$ are conjectured to have a
real part of 1/2, i.e., $z_n=1/2+i\gamma_n$.  An interesting question
is how the $\gamma_n$ are distributed on the real axis.  To this end,
it was argued that the two-point correlation function of the
$\gamma_n$ has the form $Y_2(r)=1-[\sin(\pi r)/(\pi r)]^2$
\cite{Mont73}.  This is identical to the result obtained for the
unitary ensemble of RMT and consistent with a conjecture (apparently
by Polya and Hilbert) according to which the the zeros of $\zeta(z)$
are related to the eigenvalues of a complex Hermitian operator.  By
computing the $\gamma_n$ numerically up to order $10^{20}$
\cite{Odly87}, it was shown that their distribution indeed follows the
RMT prediction for the unitary ensemble (for large enough $\gamma_n$).

\subsection*{Universal eigenvalue fluctuations in Quantum chromodynamics (QCD)} 

QCD is the theory of the {\em strong interactions}, describing the
interaction of {\em quarks} and {\em gluons} which are the basic
constituents of {\em hadrons}.  QCD is a highly complex and nonlinear
theory for which most nonperturbative results have been obtained
numerically in {\em lattice QCD} using the worlds fastest {\em
  supercomputers}. The {\em Euclidean QCD partition function} is given
by
\be
Z(m) = \int DA\, {\det}^{N_f} (D + m)e^{-S_{YM}(A)/\hbar},
\label{ZQCD}
\ee where $S_{YM}$ is the Euclidean {\em Yang-Mills action} and the
{\em path integral} is over all $SU(N_c)$ valued Yang-Mills fields
$A_\mu^{ij}$ ($\mu$ is the {\em Lorentz index}, $N_c$ the number of
colors, and $N_f$ the number of quark flavors).  The {\em Euclidean
  Dirac operator} is defined by $D=\gamma_\mu\partial_\mu+ig\gamma_\mu
A_\mu$, where $g$ is the coupling constant and $\gamma_\mu$ are the
{\em Euclidean gamma matrices}.  Because of the {\em chiral symmetry}
of QCD, in a chiral basis the matrix of $D$ has the block structure
\begin{equation}
  \label{Dirac}
  iD=\left(\matrix{0&T\cr T^\dagger&0}\right)\:.
\end{equation}
In a lattice formulation, the dimension of the matrix $T$ is a
multiple of the total number of lattice points. The smallest
eigenvalues of the Dirac operator play an important role in the QCD
partition function. In particular, the {\em order parameter} of the
{\em chiral phase transition} is given by
\be
\Sigma = \lim_{m \rightarrow 0} \lim_{V \rightarrow \infty } \frac{\pi 
\langle\rho(0)\rangle }{V},
\label{BC}
\ee
where $\langle\rho(\lambda)\rangle$ is the average spectral density of the 
Dirac operator and $V$ is the volume of space-time.

Although the QCD partition function can only be calculated
numerically, in certain domains of the parameter space it is possible
to construct {\em effective theories} which can be solved
analytically.  An important ingredient is the chiral symmetry of the
QCD Lagrangian which is broken spontaneously in the ground state.
Considering Euclidean QCD in a finite volume, the low-energy behavior
of the theory can be described in terms of an {\em effective chiral
  Lagrangian} if the linear length, $L$, of the box is much larger
than the inverse of a typical hadronic scale.  Furthermore, if $L$ is
smaller than the inverse of the mass of the pion, which is the {\em
  Goldstone boson} of chiral symmetry breaking, then the kinetic terms
in the chiral Lagrangian can be neglected. It was found by Leutwyler
and Smilga that the existence of this effective partition function
imposes constraints on the eigenvalues of the QCD Dirac operator
\cite{Leut92}. However, in order to derive the full spectrum of the
Dirac operator one needs a different effective theory defined by the
{\em partially quenched chiral Lagrangian}, which in addition to the
usual quarks includes a {\em valence quark} and its {\em superpartner}
\cite{Toublan}.  As is the case with the usual chiral Lagrangian, the
kinetic terms of this Lagrangian can be neglected if the inverse mass
of the Goldstone bosons corresponding to the valence quark mass is
much larger than the size of the box. It has been shown that in this
domain the corresponding spectral correlators are given by the chiral
ensembles which have the same block structure as the Dirac operator
(\ref{Dirac}). The $\beta$-value of the ensemble is determined by
$N_c$ and the representation of the fermions. The energy scale for the
validity of chiral RMT is the equivalent of the Thouless energy and is
given by $F^2/\Sigma L^2$, where $F$ is the {\em pion decay constant}
that enters in the chiral Lagrangian.

The fluctuation properties of the Dirac eigenvalues can be studied
directly by diagonalizing the lattice QCD Dirac operator. Correlations
in the bulk of the spectrum agree perfectly with the various RMT
results \cite{Hala95}. However, as was already pointed out, the small
Dirac eigenvalues are physically much more interesting. Because of the
relation (\ref{BC}), the spacing of the low-lying eigenvalues goes
like $1/(V\Sigma)$.  To resolve individual eigenvalues one has to
magnify the energy scale by a factor of $V\Sigma$ and consider the
microscopic spectral density \cite{Shur93} defined in (\ref{rhosu}).

Because of the chiral structure of the Dirac operator in
(\ref{Dirac}), all nonzero eigenvalues of $iD$ come in pairs
$\pm\lambda_n$, leading to level repulsion at zero.  This is reflected
in the fact that $\rho_s(0)=0$ even though $\lim_{\lambda \rightarrow
  0} \lim_{V\rightarrow \infty} \rho(\lambda)/V>0$.  The spectrum is
said to have a ``hard edge'' at $\lambda=0$.

\begin{figure}[ht]
  \centerline{\includegraphics[height=60mm]{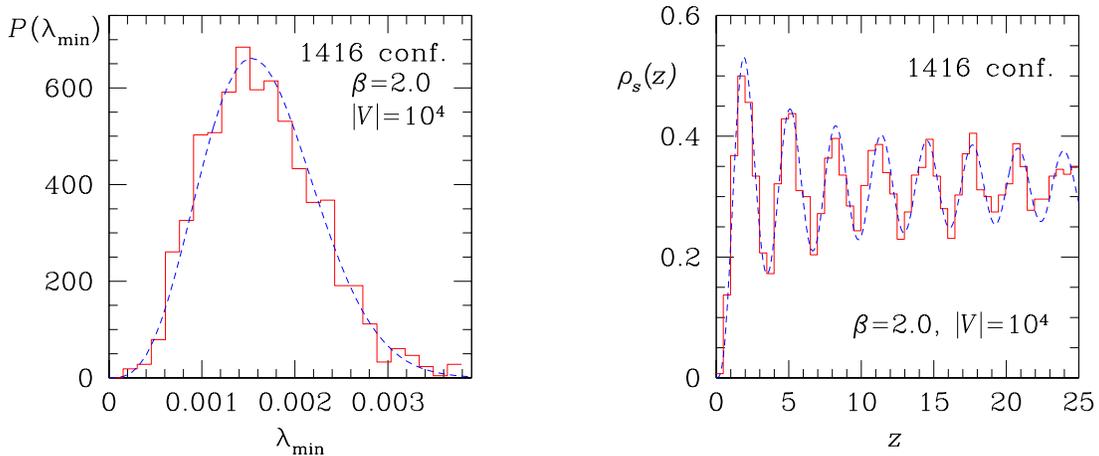}}

\caption
   { Distribution of the smallest eigenvalue (left) and
    microscopic spectral density (right) of the QCD Dirac operator.
    The histograms represent lattice data in quenched SU(2) with
    staggered fermions on a $10^4$ lattice using $\beta=4/g^2=2.0$
    (not to be confused with the Dyson index $\beta$).  The dashed
    curves are the parameter-free RMT predictions.  Taken from
    Ref.~\protect\cite{Berb98} with kind permission from the American
    Physical Society.}
\label{Fig3}
\end{figure}
The result for $\rho_s(z)$ for the chGUE (appropriate for QCD with
three and more colors) and gauge fields with
topological charge $\nu$ reads \cite{Verb93,Verb94a}
\begin{equation}
  \rho_s(z)=\frac{z}{2}\left[ J_{N_f+|\nu|}^2(z)
    -J_{N_f+|\nu|+1}(z)J_{N_f+|\nu|-1}(z)\right]\:,
\end{equation}
where $J$ denotes the Bessel function. The results for the chGSE and
the chGOE are more complicated.  Lattice QCD data agree with RMT
predictions as seen in Fig.~\ref{Fig3} which represents results
corresponding to the chGSE.

\subsection*{QCD at nonzero temperature and chemical potential}
Random-matrix models can also be used to model and analyze generic
properties of the chiral symmetry restoration phase transition at
finite temperature or finite baryon chemical potential $\mu$.  For
example, the effect of the chemical potential can be described by the
{\em non-Hermitian deformation} (\ref{qcd+mu}) of the chGUE.  The
eigenvalues of such a matrix are not constrained to lie on the real
axis.  The quantity that signals chiral symmetry breaking is the
discontinuity (a cut) of the averaged resolvent $\langle G(z) \rangle$
at $z=0$. One can calculate $\langle G(z) \rangle$ in a theory with
$n\ne0$, which corresponds to QCD with $n$ species of quarks. There is
a critical value of $\mu$ above which $\langle G(z) \rangle$ becomes
continuous at $z=0$, and, therefore, chiral symmetry is restored. In
lattice Monte Carlo the problem of calculating the partition function
and expectation values such as $\langle G(z) \rangle$ at finite $\mu$,
which are of a paramount interest to experiment, is still unresolved.
The difficulty lies in the fact that the determinant of the Dirac
matrix is complex and cannot be used as part of the probabilistic
measure to generate configurations using the Monte Carlo method.
For this reason, exploratory simulations at finite $\mu$ have only
been done in the {\em quenched approximation} in which the fermion
determinant is ignored, $n=0$.  The results of such simulations were
in a puzzling contradiction with physical expectations: The transition
to restoration of chiral symmetry occurs at $\mu=0$ in the quenched
approximation.

The chiral random matrix model at $\mu \ne 0$ 
allows for a clean analytical explanation of this
behavior, since one can calculate $\langle G(z) \rangle$ both at $n=0$
and $n\ne0$.  (As before, the number of replicas is denoted by $n$.)
The behavior of $\langle G(z) \rangle$ at $n=0$ and
$n\ne0$ is drastically different. While at $n\ne0$ the
non-analyticities of $\langle G(z) \rangle$ come in the form of
one-dimensional cuts, for $n=0$ they form two-dimensional regions,
similar to the {\em Ginibre circle} in the case of the non-Hermitian GUE.
This means that the $n=0$ (quenched) theory is not a good
approximation to the $n\ne0$ (full) theory at finite $\mu$, when the
Dirac operator is non-Hermitian.  The quenched theory is an
approximation (or, the $n\to0$ limit) of a theory with the determinant
of the Dirac operator replaced by its absolute value, which has
different properties at finite $\mu$ \cite{Step96}.

\subsection*{Quantum gravity in two dimensions.} 

In all cases we have
discussed so far, the random-matrix model was constructed for the
Hamiltonian (or a similar operator) of the system, and the universal
properties were independent of the distribution of the random matrix.
In contrast, in quantum gravity the elementary fields are replaced by
matrices, and the details of the matrix potential do influence the
results. For a recent review we refer to \cite{zinn}.

Two dimensional {\em quantum gravity} is closely related to 
{\em string theory}. 
The elementary degrees of freedom are the positions of the string in
$d$ dimensions.  The action, $S$, of the theory involves kinematic
terms and the metric.  The {\em partition function}, $Z$, is then given as a
path integral of $\exp(-S)$ over all possible positions and metrics.
The string sweeps out two-dimensional surfaces, and $Z$ can be
computed in a so-called {\em genus expansion}, i.e., as a sum over all
possible topologies of these surfaces.  This is typically done by
discretizing the surfaces.  One can then construct {\em dual graphs} by
connecting the centers of adjacent polygons (with $n$ sides).  These
dual graphs turn out to be the {\em Feynman diagrams} of a
$\varphi^n$-theory in zero dimensions which can be reformulated in
terms of a matrix model.  The partition function of this model is
given by
\begin{equation}
  Z=\int {\cal D} M\:e^{-N{\rm tr\:}v(M)} \quad {\rm with} \quad
  v(M)=\sum_{n\ge1}g_nM^n \:,
\end{equation} 
where the $M$ are Hermitian matrices of dimension $N$ and the $g_n$
are coupling constants involving appropriate powers of the
cosmological constant.  The mathematical methods used to deal with the
matrix model of quantum gravity are closely related to those employed
in RMT, giving rise to a useful interchange between the two areas.

\vspace*{0.5cm}
\noindent{\bf Acknowledgements}. We are grateful for the  
numerous discussions with collaborators and friends 
this work has benefited from. 
Peter Sellinger is acknowledged for
the {\tt potrace} software that greatly improved the quality of 
Fig. 1 and Fig. 2. This work was partially supported
by NSF grant PHY97-22101 (MAS),
U.S. DOE grant DE-FG-88ER40388 (JJMV) 
and by the Deutsche Forschungsgemeinschaft (TW).

\end{document}